\documentclass[apj,iop,twocolumn]{aastex62}

\pdfoutput=1

\hypersetup{linkcolor=red,citecolor=blue,filecolor=cyan,urlcolor=magenta}

\definecolor{caribbeangreen}{rgb}{0.0, 0.8, 0.6}

\newcommand{\Ms}{\mbox{$M_{s}$}}

\accepted{for publication in the Astronomical Journal}

\shorttitle{Circumstellar Dust Distribution in Systems with two Planets in Resonance}
\shortauthors{Marzari et al.}

\reportnum{LA-UR-18-31165}
\begin{document}

\title{\textsc{Circumstellar Dust Distribution in Systems with two Planets in Resonance}}

\author[0000-0003-0724-9987]{Francesco Marzari}
\affiliation{Department of Physics and Astronomy, University of Padova,
via Marzolo 8, 35131, Padova, Italy}
\email{francesco.marzari@pd.infn.it}

\author[0000-0002-2064-0801]{Gennaro D'Angelo}
\affiliation{Theoretical Division, Los Alamos National Laboratory, Los Alamos, NM 87545, USA}
\email{gennaro@lanl.gov}

\author[0000-0003-3754-1639]{Giovanni Picogna}
\affiliation{Universit{\"a}ts-Sternwarte M{\"u}nchen, Scheinerstra{\ss}e 1, 81679, Munich, Germany}
\email{picogna@usm.lmu.de}



\correspondingauthor{Francesco Marzari}

\begin{abstract}

We investigate via numerical modeling the effects of two planets locked
in resonance, and migrating outward, on the dust distribution of the natal
circumstellar disk. We aim to test whether the dust distribution
exhibits peculiar features arising from the interplay among the gravitational
perturbations of the planets in resonance, the evolution of the gas, and its influence
on the dust grains' dynamics.
We focus on the 3:2 and 2:1 resonance, where the trapping may be caused by the
convergent migration of a Jupiter- and Saturn-mass planet, preceding the common gap formation and ensuing outward (or inward) migration.
Models show that a common gap also forms in the dust component -- similarly to what
a single, more massive planet would generate -- and that outward migration leads to a progressive widening of the dust gap and to a decoupling from the gas gap.
As the system evolves, a significantly wider gap is observed in the dust distribution, which ceases to overlap with the gas gap in the inner disk regions.
At the outer edge of the gas gap, outward migration of the planets produces an
over-density of dust particles, which evolve differently in the 3:2 and 2:1 resonances.
For the 3:2, the dust trap at the gap's outer edge is partly efficient and a significant
fraction of the grains filters through the gap. For the 2:1 resonance, the trap is more
efficient and very few grains cross the gap, while the vast majority accumulate at the outer edge of the gap.


\end{abstract}

\keywords{planets and satellites: detection --- planets and satellites: dynamical evolution and stability ---
planets and satellites: gaseous planets ---
protoplanetary disks --- planet-disk interactions}

%

\section{Introduction} \label{sec:intro}

\defcitealias{dangelo2012}{DM12}

In the early stages of evolution of a planetary system, dust, gas 
and planets may coexist. By probing the dust content of protoplanetary 
disks, ALMA and SPHERE observations of dust continuum emission have 
revealed the presence of gaps, rings, and spiral structures 
\citep[e.g.,][]{andrews2016,isella2016,feldt2017}, 
which may represent signatures of planet-dust interactions. 
Small planets, not massive enough to carve a gap in the gas, may still 
be able to open gaps in the dust distribution due to their tidal torque. 
Exterior to the planet's orbit, the gravitational torque can counterbalance 
the aerodynamic drag torque, halting the inward drift of the dust.
Interior to the planet's orbit, these torques add up and push the dust 
away, toward the star. Numerical and analytical modeling 
\citep[e.g.,][]{paamel2004,paamel2006,rice2006,zhu2014,rosotti2016,dipierro2017,ricci2018,picogna2018}
explored this type of interactions and showed that planets with masses as low as  $\approx 5\,M_{\oplus}$ (depending on the gas temperature and level 
of turbulence viscosity) can produce detectable signatures in a disk. The solid component of the 
disk has been modeled either as a dust fluid or using a large number of Lagrangian particles 
representative of the dust dynamics. 


Characteristic features are more easily produced by the planet's tidal 
torque in the dust, rather than in the gas, distribution because the 
disk's gas is also subjected to a viscous torque, which tends to smooth 
out large density gradients \citep{pringle1981}.
Planets massive enough to perturb the gas and open a gap in the gas 
distribution can create a pressure maxima at the outer gap edge, 
where dust particles can be trapped and pile up, enhancing the local 
density of the solids and producing a gap in the dust distribution 
around the planet's orbit. If dust is completely unable to filter 
through the gap, a cavity in the dust distribution develops.
Depending on the disk's physical properties, already for masses larger 
than about $20\,M_{\oplus}$ \citep[e.g.,][]{lambrechts2014,bit2018}, 
a planet may be able to halt the inward flux of small solids and 
produce a cavity, observationally defined as a transition disk.
According to \citep{rice2006}, for Jupiter--like planets the filtration process is size 
dependent and dust grains typically  smaller than 10 $\mu$m may cross 
the planet gap and reach the inner region of the disk increasing the
 small particles population in the inner region of the disk.
Because of their observational signatures, dust features can provide 
invaluable insights into the physics of planet-dust interactions and 
into the architecture of planetary systems.

In this paper, we focus on two giant planets, orbiting in a gaseous disk and locked in mean motion resonance 
(MMR), and explore the type of signatures they can impose on the dust 
distribution. 
A significant number of exoplanets are close to low-order resonances, 
like the 2:1 or 3:2 \citep{wright2011,fab2012}. 
HD~45364 \citep{rein2010}, KOI~55 (Kepler~70) \citep{charpinet2011} 
and other Kepler systems \citep{steffen2013} are suspected to be in 
the 3:2 resonance while Gliese~876, HD~82943 and HD~37124 \citep{wright2011} 
are probably in the 2:1 resonance. The K2-32 system hosts three planets 
near 3:2:1 resonances \citep{petigura2017}. 
Many other systems may have been in resonance during their migration 
epoch, escaping from it at later times due to gas dissipation \citep{marza2018}
and/or to
gravitational interactions with leftover planetesimals \citep{chat2015}. 

A resonant configuration may be either the outcome of the formation 
process or be a consequence of the early dynamical evolution of 
a planetary system. The formation of the first planet may trigger 
the growth of additional massive bodies in resonance, creating 
a resonance chain, like in Kepler~223 \citep{tan2014,mills2016}. 
A dynamical mechanism that may lead to resonant trapping of planets 
is convergent migration. If the outer planet migrates inward faster 
than does the outer planet \citep{masset2001,Leepeale2002}, they eventually 
become trapped in a mean motion resonance. The faster migration of the 
outer planet may be due, e.g., to its mass being lower than that of 
the inner planet. In the case of a Jupiter/Saturn-type pair, the 
exterior planet is unable to fully empty its co-orbital region and 
its migration speed is similar to type~I migration driven by corotation
and Lindblad resonances \citep[e.g.,][]{tanaka2002}. The inner planet, 
with a Jupiter mass or higher, will instead open a gap and drift 
inward at the typically slower rate of type~II migration 
\citep[e.g.,][and references therein]{dangelo2008,ragusa2018}. 
The outer planet will approach the inner one and can be captured in 
a mean motion resonance, typically either the 2:1, the first to be 
encountered, or the 3:2 (more compact resonances are possible, but
they require larger migration velocities during the approach phase). Once trapped 
in resonance, both planets may contribute to open a common gap 
possibly resulting in a regime of coupled migration. 

The shape and depth of the common gap and the torques exerted 
on the pair of planets strongly depend on the type of resonance 
(either the 3:2 or the 2:1 MMR) and on the disk parameters 
\citep{masset2001,lee2002,adams2005,thommes2005,beauge2006,crida2008,dangelo2012}. 
Inward or outward migration may result, which is determined by 
the values of viscosity in the disk, the density and temperature 
profiles of the gas, and the masses of the two planets. 
Once captured in the mean motion resonance, the period of coupled 
inward/outward migration may continue until the disk eventually 
dissipates (assuming no interactions with other possible companions
in the system). 
The ``grand tack'' scenario \citep{walsh2011} is a model based on 
this process. It was invoked for explaining some features of the 
solar system, like the small mass of Mars and the compositional 
differences in the asteroid belt. 
It proposes that Saturn during its inward migration was trapped in 
either a 3:2 or 2:1 MMR with Jupiter 
\citep{morby2007,pierens2008,dangelo2012,pierens2014}. 
The formation of a common gap in the gas leads to a reversal of the 
torque sign, causing the interior planet to migrate outward, carrying the outer planet with it (via resonant forcing). The locked outward migration 
of the two planets continues as long as an appropriate torque imbalance 
can be sustained. 

In this paper we explore the evolution of dust particles in a circumstellar 
disk where two planets are trapped in resonance and migrate outward. 
We look for peculiar features in the dust distribution, related to this 
dynamical configuration, which might give rise to observable signatures 
in high resolution images of disks acquired, 
e.g., with ALMA and SPHERE. 
The shape of the gap of two planets 
in resonance differs from that carved by a single one and the dust trapping 
at the outer border may change depending on the type of resonance. 
In addition, the outward migration of the planet pair may lead to 
the formation of morphological traits in the disk that can be detected 
in images and interpreted as due to two resonant planets. Two-dimensional
(2D) hydrodynamical simulations of the outward migration of two planets 
in resonance, either in the 3:2 or 2:1 MMR, are performed and the trajectories 
of solid particles of various sizes are integrated under the effect of 
drag forces induced by their interaction with the gas. 
In Section~\ref{sec:resonances}, we briefly summarize the dynamics of 
two planets embedded in a disk with their orbits locked in a MMR. 
In Section~\ref{sec:numerical}, we outline the numerical model and the 
initial setup of the simulations. In Section~\ref{sec:32mmr}, we describe 
the evolution of the dust in the proximity of the common gap developed 
by a 3:2 MMR while, in Section~\ref{sec:21mmr}, we focus on the 2:1 MMR. Section~\ref{sec:end} is dedicated to the discussion of 
our results and conclusions.

\section{Dynamical Evolution in the 2:1 and 3:2 Resonances}
\label{sec:resonances}

The most common scenario for the assembly of a 2:1 or 3:2 MMR is that 
the differential (convergent) orbital migration of two planets drives 
them into the resonance. If the resonant forcing outweighs the 
tidal forcing, the orbits can become locked. Otherwise, convergent 
migration continues until either another resonance is encountered or 
dynamical instability ensues \citep{marza2010}.
Since the planets orbit in a relatively compact configuration, the local 
thermodynamics conditions are similar in a smooth disk and, therefore, 
differential orbital migration typically requires that planets have 
different masses. 
Convergent migration also places some constraints on the disk's properties.

Analytical arguments \citep{quillen2006,mustill2011} suggest that 
a less massive planet, of mass $M_{2}$, can be locked into the 2:1 orbital 
resonance with an interior and more massive planet, of mass $M_{1}$, if 
$|\dot{a}_{2}-\dot{a_{1}}|\lesssim 1.2(M_{1}/\Ms)^{4/3}a_{2}\Omega_{2}$,
where $a_{1}$ and $a_{2}$ are the planets' semi-major axes, and $\Omega_{2}$ 
is the orbital frequency of the outer planet. Assuming that orbital 
migration is dominated by the \textit{type~I} 
\citep[or a modified type~I,][]{dangelo2008,dangelo2010}
migration mode, \citetalias{dangelo2012} found that the 1:2 resonance 
capture occurs when
\begin{equation}
\left(\frac{a^{2}\Sigma}{\Ms}\right)\lesssim\left(\frac{H}{a}\right)^{2}%
\left(\frac{\Ms}{M_{2}}\right)\left(\frac{M_{1}}{\Ms}\right)^{4/3},
\label{eq:C21}
\end{equation}
where $\Sigma$ and $H$ are the surface density and pressure  scale-height 
of the disk at around the outer planet's location. The predictions of 
the inequality above agree with results from direct hydrodynamical 
calculations of 2:1 resonance capture. 
For a given gas surface density, locking in the 2:1 MMR is more likely 
in warm disks, because of the slower (relative) migration velocity.
If the inequality~(\ref{eq:C21}) is not satisfied, the tidal forcing 
can break the resonance and another, more compact, resonant configuration 
may be established.

According to condition~(\ref{eq:C21}), the gas surface density required 
for capture in the 2:1 resonance is $\Sigma\lesssim 100\,\mathrm{g\,cm^{-2}}$ 
in the $5\,\mathrm{au}$ region, when $H/a\approx 0.03$. 
For typical circumstellar disks, these densities are obtained after 
$\sim 0.5\,\mathrm{Myr}$ of evolution 
\citepalias[see also \citeauthor{dalessio2001} \citeyear{dalessio2001}]{dangelo2012}, 
long before a Jupiter-mass planet can form
\citep[e.g.][]{lissauer2009,movshovitz2010,dangelo2014}.

One critical aspect of the 2:1 resonance is that the continued locked 
migration of the planets can lead to rapid growth of their orbital 
eccentricities \citep{lee2002,lee2003,ferraz2003,lee2004,teyssandier2014}. 
If the mass ratio $M_{1}/M_{2}$ is similar to that of Jupiter and 
Saturn, i.e. $\approx 3.4$, then the outer planet achieves a higher 
eccentricity. 
This was shown by \cite{lee2004} for various mass ratios of the planets. 
The peak values of eccentricity reached by the two planets depend on 
the damping effect caused by the gravitational interaction with the 
disk \citep{teyssandier2014}. This outcome provides conditions 
on the disk density around the orbits of the two planets. 
Different eccentricities of the planets can lead to different gap shapes,
which may significantly affect the dust density distribution in the 
disk where the gas eccentric perturbations are significant. 

\section{The Numerical Models} 
\label{sec:numerical}

The evolution of a pair of planets interacting with a gaseous disk 
is simulated by means of the 2D FARGO code \citep{Masset}, as modified by \cite{mullerkley2012}, which 
solves the Navier-Stokes equations for the disk on a polar grid 
$(r,\phi)$. 
We perform simulations where the energy equation includes viscous 
heating and radiative cooling through the disk surface
\begin{equation}
\frac{\partial E}{\partial t} + \nabla \cdot (E\,\mathbf{u})
                =  -P \, \nabla\cdot\mathbf{u} + Q^{+} - Q^{-},
\label{eq:eneq}
\end{equation}
in which $E$ and $P$ are the total energy (surface) density and 
pressure, respectively, and $\mathbf{u}$ is the gas velocity field.
In Equation~(\ref{eq:eneq}), $Q^{+}$ is the viscous dissipation 
term, computed from the components of the viscosity stress tensor 
\citep{m&m}, 
which includes a physical kinematic viscosity and a 
von Neumann-Richtmyer artificial (bulk) viscosity,
implemented as in \cite{stone92} and where the
coefficient defining the number of zones 
over which shock fronts are smeared out is set 
to $1.41$. 
The term 
$Q^{-} = 2 \sigma_{\mathrm{SB}} T_{\mathrm{eff}}^4$ 
is the local radiative cooling. The effective
temperature $T_{\mathrm{eff}}$ is computed by 
using the \cite{linbell} formulae for the Rosseland
mean opacity, $\kappa$ ($\sigma_{\mathrm{SB}}$ is
the Stefan-Boltzmann constant). 

The trajectories of small dust grains with sizes 
$10 \,\mathrm{\mu m}$, 
$100 \,\mathrm{\mu m}$, 
$1 \,\mathrm{mm}$ and 
$1 \,\mathrm{cm}$,   
for a total of $4\times 10^{5}$ particles, are integrated along 
with the hydrodynamical evolution of the gas. 
Although the largest-size particles do not strictly qualify 
as dust grains in the typical context of the interstellar medium 
\citep{draine2011}, we model them because of the their importance
in the context of protoplanetary disks and refer to them as grains 
for simplicity.
These grain sizes represent typical values that can be detected 
by ALMA in the infrared.
The aerodynamic forces acting on the dust particles are computed 
as in \cite{picogna2015}. 
The drag force acting on a spherical dust particle of radius $s$, moving 
with a velocity $\mathbf{v}$ relative to the gas 
is given by
\begin{equation}
\mathbf{F}_D = (1-f) \mathbf{F}_{D,\mathrm{E}} + f \mathbf F_{D,\mathrm{S}}
\label{eq:FD}
\end{equation} 
where 
\begin{eqnarray}
\mathbf{F}_{D,\mathrm{E}} & = & -\frac{4}{3}\pi s^{2} \rho_{g} v_\mathrm{th} \mathbf{v} \label{eq:FDE}\\
\mathbf{F}_{D,\mathrm{S}} & = & -\frac{1}{2}\pi s^{2} C_D \rho_{g} v \mathbf{v}
\label{eq:FDS}
\end{eqnarray}
are drag forces in the Epstein and Stokes regimes. In Equation~(\ref{eq:FDE}),
$v_\mathrm{th} = \sqrt{8 k_{\mathrm{B}} T/(\pi \mu m_\mathrm{H}}) $ is the mean 
thermal velocity of the gas molecules, $T$ the local temperature of the 
gas, $\rho_{g}$ the gas volume density, $m_\mathrm{H}$ the hydrogen atom 
mass, and $\mu$ the mean molecular weight of 
the gas. The drag force in the Stokes 
regime is proportional to the drag coefficient, $C_D$, whose value is
taken from \citet{weidenschilling1977} and depends on the Reynolds number.
The transition between the two drag forces is determined by the coefficient 
$f$, given by 
\begin{equation}
f = {s \over {s + \lambda} } = {1 \over {1 + \mathrm{Kn}}},
\end{equation}
where $\lambda$ is the mean free path of the gas molecules and $\mathrm{Kn} = \lambda/s$ is the Knudsen number. Comparing the 
expressions for $\mathbf{F}_{D,\mathrm{E}}$ and $\mathbf{F}_{D,\mathrm{S}}$, 
it can be shown that they are equal when $\mathrm{Kn} = 4/9$ for Reynolds 
numbers $<1$ \citep[see][]{weidenschilling1977}. 
Due to drag forces, particles experience a radial drift relative to 
the gas and they also respond to density and velocity gradients
in the gas. The drift velocity (relative to the gas), in conditions of 
stationary gas surface density, can be approximated as 
\citep{birnstiel2010,pinilla2012}
\begin{equation}
v_\mathrm{drift} = \frac{1}{\mathrm{St}^{-1} + \mathrm{St}}%
                   \left(\frac{\partial P}{\partial r}\right)%
                   \frac{1}{ \rho_g \Omega}.
\end{equation}
Indicating with $m_{s}$ the mass of a particle,
$\mathrm{St}=m_{s} v \Omega/\mathbf{F}_{D}$ represents the Stokes number 
(sometimes referred to as non-dimensional stopping time) and $\Omega$ 
the Keplerian frequency of the disk. Since $v_\mathrm{drift}$ depends 
on the radial derivative of the gas pressure $P$, any local pressure 
maximum in the disk (with a significant azimuthal extent) will collect 
and trap grains, both orbiting in the vicinity of the maximum and those 
drifting inward from the outer disk regions. 
Under barotropic conditions, a local pressure
maximum translates into a density maximum that
generates locations where dust grains can be
confined and, as a result, where their density may
increase. The common gap formed by two resonant planets can 
produce such confinement locations, in particular at its outer border. 

Additional calculations are performed with the code employed by 
\citetalias{dangelo2012}, which apply an orbital advection 
scheme implemented along the lines of the algorithm described by 
\citet[][]{Masset}. Further details are given in \citetalias{dangelo2012}. 
The thermodynamical evolution of the solids interacting with the gas, 
the planets, and the star, is calculated using the methods and algorithms 
described in \citet{dangelo2015}. The code computes the motion of solids 
through the gas and their thermal state in a self-consistent fashion, 
including ablation and break-up. The small solids considered here are 
basically isothermal and in thermal equilibrium with the gas. Since they 
are typically very well-coupled to the gas, frictional heating is small 
and their thermal budget is mainly dictated by heating in the gas 
thermal field and by radiative cooling. Due to the low velocities 
relative to the gas, the dynamical pressure exerted by the gas is 
typically not sufficient to break up the solids. Since the particles 
modeled here are made of silicates and their temperature does not 
exceed a few hundred kelvin degrees, ablation is negligible.
In some test cases under the same initial
conditions the two codes provided very similar
results. Therefore, we decided to use the FARGO
code for most of the computations discussed herein.

\subsection{The disk setup}

The shape and depth of the common gap developed by a pair of planets 
and the outward (inward) migration rate are strongly dependent on 
the values of disk viscosity, density and temperature profiles of 
the gas, and masses of the two planets \citep{dangelo2012,masset2001}. 
Performing a full exploration of all these parameters, along with 
the study of the dust evolution, is outside the scope of this work. 
We rather focus on models in which the planets
become locked in either the 3:2 or 2:1 MMR and
migrate outward.
The investigation of the dust evolution in these 
models is performed with the goal of highlighting
some effects that may represent a common outcome 
of resonant pairs undergoing outward migration.  
Although some calculations resulted in inward
migration after resonance locking of the planet
pair, here we only report on the effects of 
outward migration on the dust particle
distribution.

To model the circumstellar disk with FARGO, we
adopt a polar, $(r,\phi)$, grid with 
$720\times 360$ elements uniformly covering 
the disk, which extends from 
$r = 0.5\,\mathrm{au}$ to $15\,\mathrm{au}$ 
from the star. The initial gas density profile is
given as 
\begin{equation}
\Sigma (r) = \Sigma_0 \left(\frac{r}{\mathrm{au}}\right)^{-p} \,\mathrm{g\,cm^{-2}}.
\label{eq:Sigma_r}
\end{equation}

When modeling the 3:2 MMR, we consider two density values, 
$\Sigma_0 = 750\,\mathrm{g\,cm^{-2}}$ and 
$\Sigma_0 = 1000\,\mathrm{g\,cm^{-2}}$, and the slope of the 
density distribution (i.e., the power index) is set to $p=1$. For the 2:1 MMR, we consider 
instead only $\Sigma_0 = 750\,\mathrm{g\,cm^{-2}}$ and $p=3/2$.  
These parameters are chosen (based on several experiments) so to 
ensure a sustained outward migration of the two planets once 
they are locked in resonance.
It is noteworthy that for smaller values of $\Sigma_0$,
i.e., $\Sigma_0 < 500 \,\mathrm{g\,cm^{-2}}$, capture is always 
in the 2:1 MMR and the migration is directed 
inward regardless of the power index $p$ 
(that was applied).

The values adopted for $\Sigma_0$ are smaller than those predicted 
for the Minimum Mass Solar Nebula \citep{weidenschilling1977,hayashi81}
since they are presumed to take into account the time required by 
the planets to grow. During this time the disk is partially dissipated 
by the viscous evolution and, possibly, 
by photo-evaporation. Nonetheless, 
these values correspond to initially relatively massive protoplanetary disk. 
In fact, allowing $0.5$--$1\,\mathrm{Myr}$ for the two planets to
form and $3$--$4\,\mathrm{Myr}$ for the disk's gas to disperse, 
the initial disk mass can be as large as $\sim 0.1\,M_{\odot}$
\citepalias{dangelo2012}.

The disk aspect ratio is initialized to $h = H/r = 0.02$. 
The shear kinematic viscosity is $\nu = \alpha H^{2} \Omega$, 
where the turbulence parameter is a constant, $\alpha=10^{-4}$ 
or $10^{-3}$ for the 3:2 MMR and $\alpha=10^{-4}$ for the 2:1 MMR.
This initial setup is similar to that adopted by \citet{pierens2014},
in which the authors study the outward migration of Jupiter and 
Saturn in cold disks. 
 When the planets are fully formed, the host disk is possibly evolved and may be characterized by a low viscosity. Both these factors may lead to a disk with a small aspect ratio ($h \approx 0.02$). Regardless, in the present case, this small value is required to drive outward migration when the planets are trapped in a 2:1 resonance.  Warmer, more viscous disks do not typically induce outward migration of the pair of planets, but rather inward migration
\citepalias{dangelo2012}. This condition on the aspect ratio can be relaxed for the 3:2 resonance which may drive outward migration even for large aspect ratios ($h=0.05$) \citep{masset2001,dangelo2012}.


Note that since the gas temperature evolves according to 
Equation~(\ref{eq:eneq}) and 
$H\Omega=\sqrt{k_{\mathrm{B}} T/(\mu m_\mathrm{H}})$,
the kinematic viscosity, $\nu$, is proportional to 
$T$ and therefore varies with time, both 
globally and locally around the planets.

Convergent migration determines which resonance is established.
According to condition~\ref{eq:C21}, in cold disks it is more 
likely to overcome the 2:1 resonance locking. In an unperturbed 
and stationary disk the energy balance in Equation~(\ref{eq:eneq})
reduces to 
\begin{equation}
Q^{+}\approx Q^{-},
\label{eq:eneq_sim}
\end{equation}
where viscous heating can be approximated to 
$Q^{+}\approx 9/4 \Sigma\nu\Omega^{2}$ 
and radiative cooling is 
$Q^{-} \approx 32 \sigma_{\mathrm{SB}} T^4/(3\kappa\Sigma)$
(assuming that the optical thickness to the disk midplane is 
$\kappa\Sigma/2$). At low temperatures ($T\lesssim 200$), dust 
opacity is dominated by ice grains and $\kappa\approx\kappa_{0}T^{2}$ 
\citep[e.g.,][]{linbell}. 
Therefore, $T\propto \alpha\Sigma^{2}\Omega$ and 
$H^{2}\propto \alpha\Sigma^{2}/\Omega$, which 
entails the possibility of local variations of the
pressure scale-height in the proximity of the
planets, possibly affecting the dynamics of both
the planets and the dust.
Note that the situation is quite different in a locally isothermal 
disk, in which $H$ depends on $r$ but remains constant in time and 
does not respond to local perturbations of the gas exerted by 
the planets.

\begin{figure}[hpt]
   \begin{center}
   \includegraphics[width=\hsize]{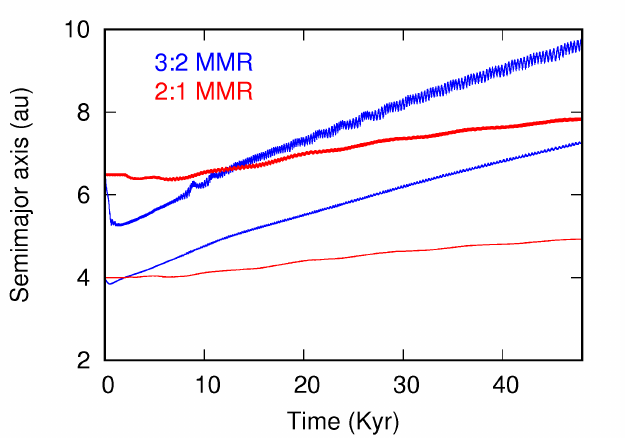}
   \includegraphics[width=\hsize]{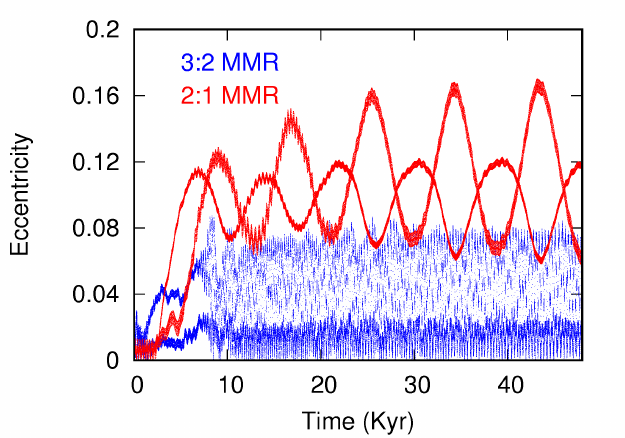}
   \end{center}
   \caption{\label{migra} Time evolution of the
   semi-major axis (top) and eccentricity of the
   two planets locked in a 3:2 (blue lines) 
   and 2:1 MMR (red line). The gas density 
   profile declines as $r^{-1}$ in the first 
   case and as $r^{-3/2}$ in the second 
   case. The migration in the 3:2 MMR case is
   faster also due to the larger density around 
   the planets' locations.}
\end{figure}
Figure~\ref{migra} shows the semi-major axis and eccentricity of the 
inner and outer planets in the two different cases. The outcome of 
the resonance locking (top panel) is in general
agreement with the inequality~(\ref{eq:C21}), 
which predicts densities somewhat lower than
$40\,\mathrm{\mathrm{g cm^{-2}}}$ at 
$r\approx 6\,\mathrm{au}$ for capture in the 2:1 
resonance. To within factors of order unity, the
agreement is reasonable also for the case with
density slope $p=3/2$ (red lines).

When the planets are locked in the 3:2 resonance (blue lines) their 
eccentricities remain low during the outward migration and that 
of the inner planet stabilizes around $0.01$ while that of the 
outer planet is on average $\approx 0.05$. In the case of the 
2:1 MMR, where the slope of $\Sigma$ is $p=3/2$, the eccentricities 
of the two planets grow on a short timescale to $0.09$ and 
$\approx 0.12$ for the inner and outer planet, respectively, 
and evolve through cycles of large anti-correlated oscillations.

\begin{figure}[hpt]
   \begin{center}
   \includegraphics[width=\hsize]{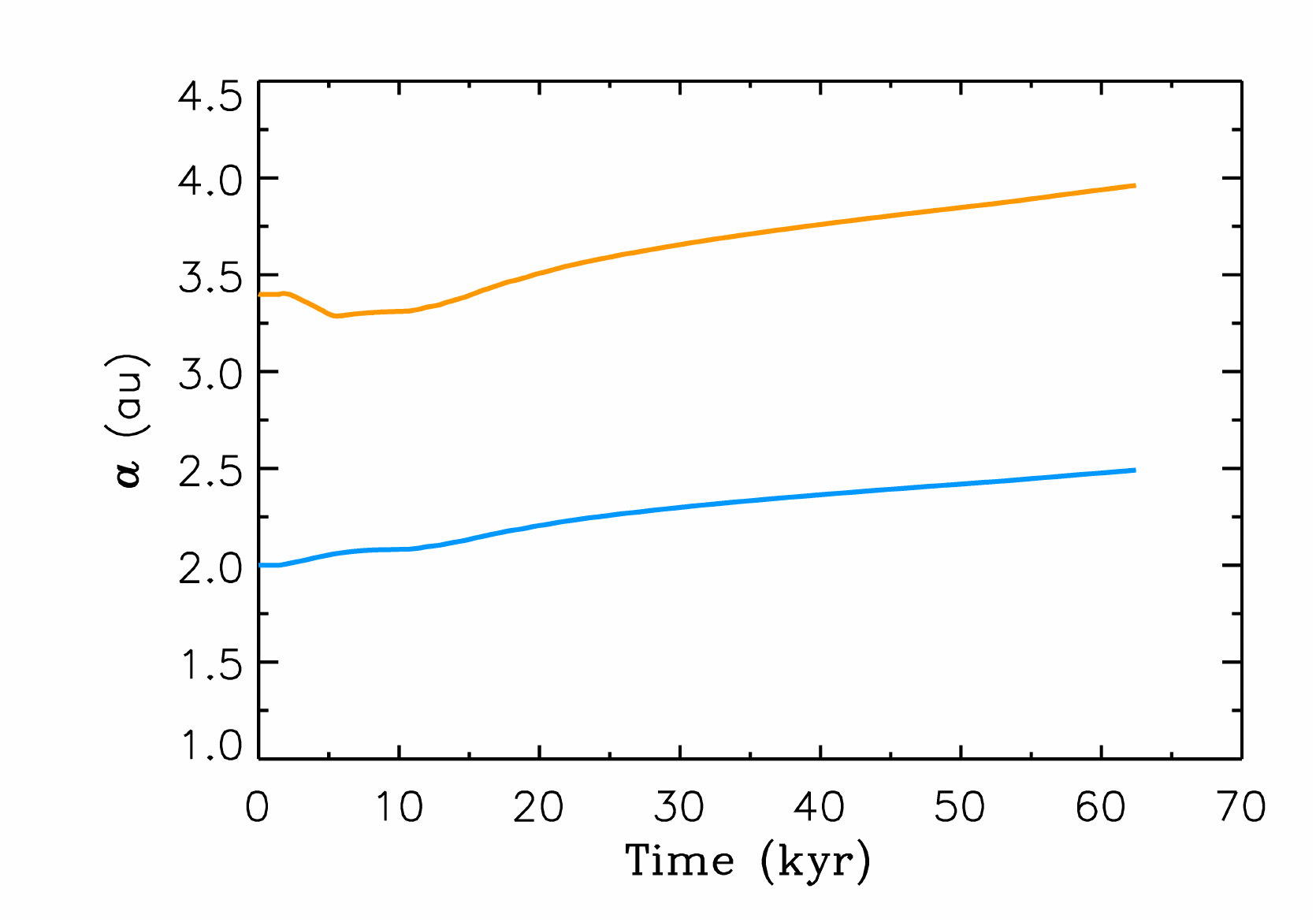}
   \includegraphics[width=\hsize]{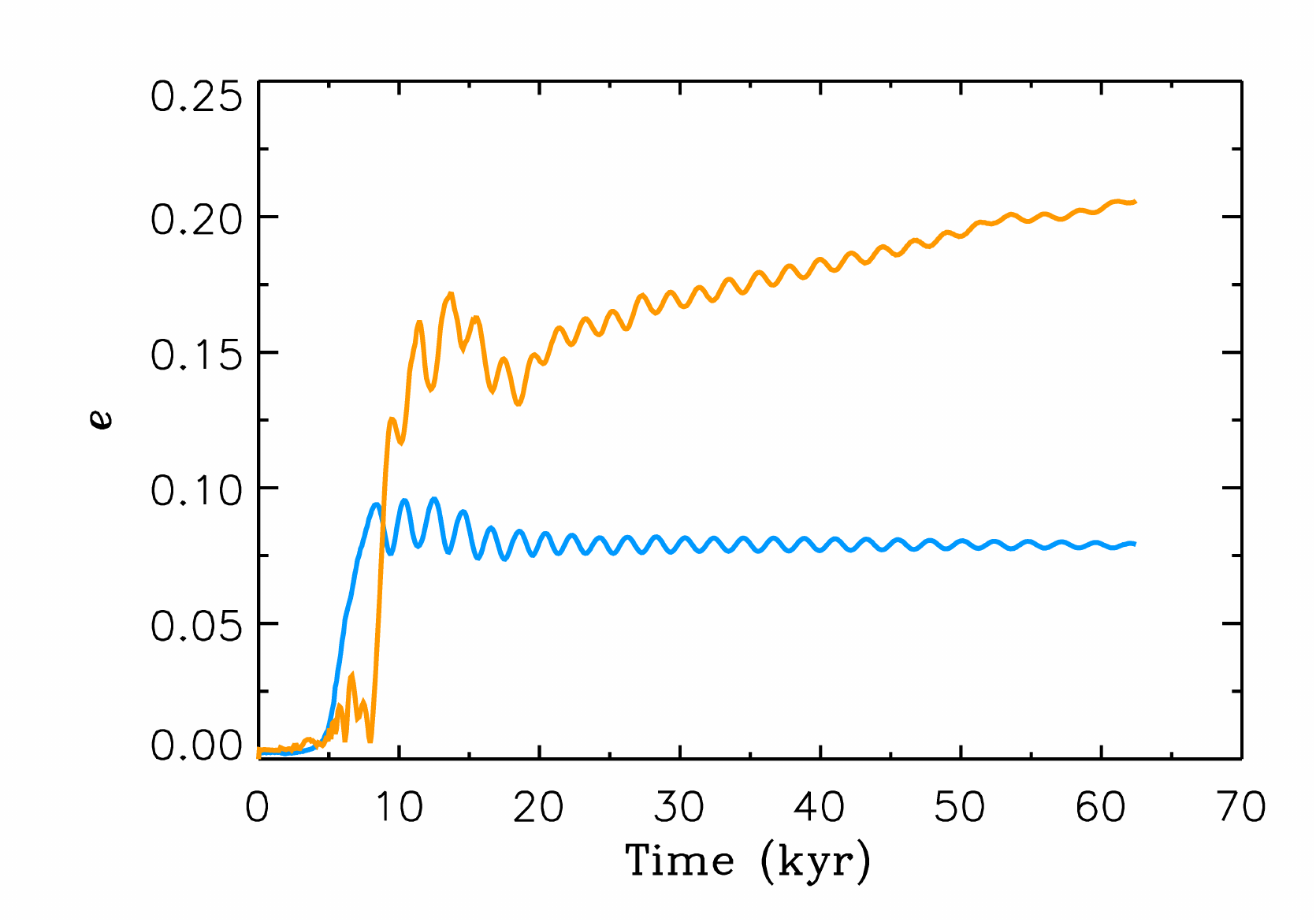}
   \end{center}
   \caption{Semi-major axis (top) and orbital
   eccentricity (bottom) evolution of a 
   Jupiter-Saturn mass pair converging into the 
   2:1 MMR and migrating outward thereafter.
   The more massive planet is on the interior
   orbit. This isothermal simulation is perform with the
   second code discussed in
   Section~\ref{sec:numerical}.}
   \label{ae21}
\end{figure}

To test the robustness of the migration scenario,
we performed an additional calculation carried out
with the other code discussed in 
Section~\ref{sec:numerical} and the results are
illustrated in Figure~\ref{ae21}. In this model,
the disk is flared and locally isothermal with 
$H/r=0.024 \,(r/\mathrm{au})^{2/7}$,
the kinematic viscosity is $\nu = 4\times 10^{-8}$
in units of $\Omega\,r^{2}$ at $r=1\,\mathrm{au}$,
corresponding to $\alpha=10^{-4}$ at
$1\,\mathrm{au}$. 
Referring to Equation~(\ref{eq:Sigma_r}), the gas
surface density is such that 
$\Sigma_0 = 530\,\mathrm{g\,cm^{-2}}$ and $p=3/2$. 
The disk extends radially from $0.25$ 
to $54\,\mathrm{au}$, with a resolution 
$\Delta r=\Delta\phi/r=0.01$. 
As expected from condition~(\ref{eq:C21}),
convergent migration drives 
the pair of planets in the 2:1 MMR (top panel). 
After the resonance is established, gap overlap
drives the pair outward. 
Similarly to the cases shown in Figure~\ref{migra},
the inner
planet acquires an orbital eccentricity $\approx 0.08$, whereas
the eccentricity of the outer planet increases beyond $0.2$
(bottom panel).
These results are overall consistent with those discussed above,
implying that the details of the disk structure may not be
critical to the outcome, as long as the gas is \emph{cold} and 
has \emph{low viscosity}.

Note that the interior, Jupiter-mass planet in
Figure~\ref{ae21} starts to migrate outward when
``released'' (tidal torques start acting on the
planets only after $1500$ years from the beginning
of the simulation, to allow for an initial 
relaxation period), prior to the onset of the 
resonance locking. 
The reason for this behavior is probably related 
to tidal interactions with a cold and low-viscosity
gas, which induce eccentric perturbations in the
disk that result in a positive torque exerted on
the planet \citep{dangelo2006}.
A similar behavior can also be seen in the top
panel of Figure~\ref{migra} (red lines).

The eccentricity evolution in the 2:1 resonance
displayed in the bottom panel of Figure~\ref{migra}
also shows large-amplitude oscillations that are
not seen in the evolution illustrated in the 
bottom panel of Figure~\ref{ae21}. This is likely
due to the local-isothermal approximation used in
the latter model. The disk is warmer and the
kinematic viscosity is not affected by disk
temperature. This is not the case in the
models of Figure~\ref{migra}, which tend to have
lower aspect ratios around the orbits of the
planets and the kinematic viscosity depends on 
the disk temperature. The smaller values
of $H/r$ and $\nu$ tend to enhance the eccentricity
of the disk's gas \citep[both in strength and
spatial domain][]{dangelo2006}. 
The exchange of eccentricity between the planet's
orbit and the disk, compounded with the resonant
forcing between the planets, may be responsible for
the resulting variations in the planet's
eccentricity evolution.

\section{The 3:2 resonance}
\label{sec:32mmr}

By applying an initial surface density with power index $p=1$ and 
$\Sigma_0 = 750\,\mathrm{g cm^{-2}}$, the two planets become 
trapped in a 3:2 MMR, and migrate outward thereafter.
A common gap is formed in the gas distribution
as a result of the tidal interaction between 
the planets and the disk. The common gap moves
along with the planets while their orbital
eccentricities remain small (see
Figure~\ref{migra}).

\begin{figure}[hpt]
\centering
\includegraphics[width=1.0 \hsize]{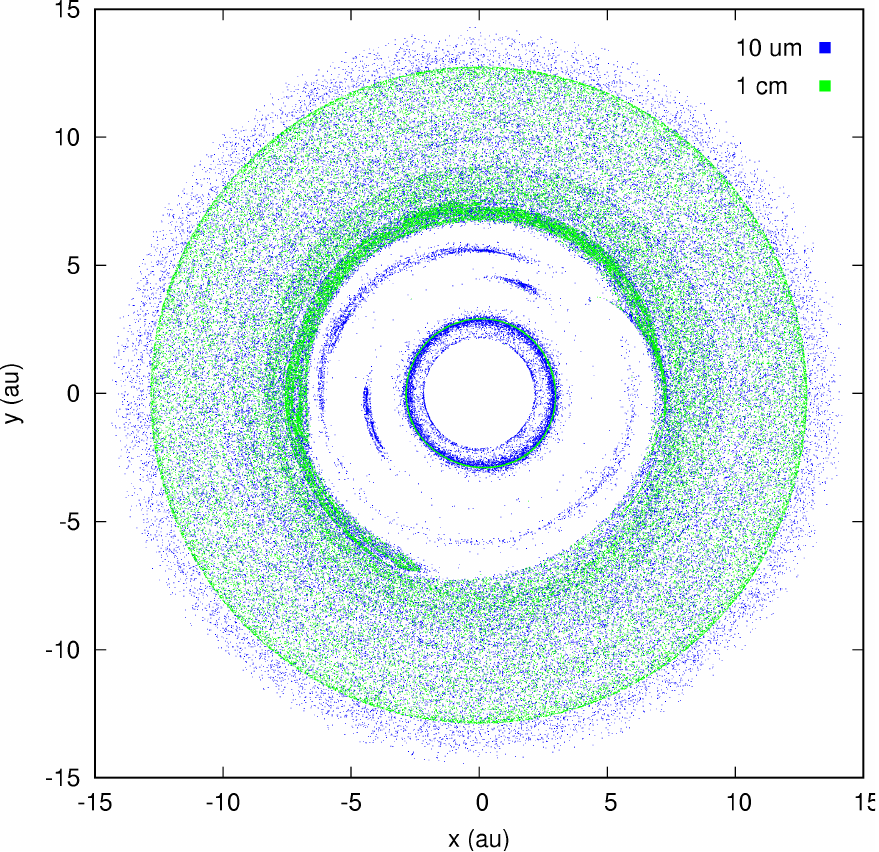}
\caption{\label{fig_00}
Dust distribution after $20\,\mathrm{kyr}$ 
of evolution. A gap forms in the dust distribution
and some dust is trapped in the coorbital regions
of the planets. The particles shown in the plot
have radii of $10\,\mu\mathrm{m}$ and
$1\,\mathrm{cm}$ and Stokes numbers of $10^{-5}$
and $10^{-2}$, respectively. While particles with
size from $10 \,\mathrm{\mu m}$ to 
$1\,\mathrm{mm}$ show a similar evolution, 
$1 \,\mathrm{cm}$ size grains exhibit some
morphological differences in the spatial
distribution compared to the smaller ones. }
\end{figure}

To test the formation of a common gap also in the dust distribution, 
we first run a simulation in which the planet masses grow on a timescale 
of $10^{3}$~yr, until they attain their final values.
The dust particles are introduced in the simulation from the beginning
and they evolve under the influence of gravitational
and drag forces, and are affected by the formation of the gas gap 
during and after capture in resonance. 
The initial dust distribution extends from $2$ to $15\,\mathrm{au}$ 
from the star and all dust grains migrating inside of $r=2 \,\mathrm{au}$
are discarded from the simulation so to reduce the
computing time.
Figure~\ref{fig_00} shows dust spatial distributions after 
$20\,\mathrm{kyr}$. A gap in the dust distribution forms, for all 
particle sizes considered here (from $10 \,\mathrm{\mu m}$ to 
$1 \,\mathrm{cm}$), overlapping the gas gap created by the pair 
of planets in MMR. The dust gap is not completely empty since some 
grains are trapped in the coorbital regions of both orbits, possibly 
because of the assumed fast growth of the planet masses 
\citep{marscho98}. 

In Figure~\ref{fig_00}, we only plot the spatial distributions of 
$10 \,\mathrm{\mu m}$ and $1 \,\mathrm{cm}$ size particles. 
In fact, while grains in the size range between $10 \,\mathrm{\mu m}$ 
and $1 \,\mathrm{mm}$ show a similar evolutionary behavior in terms
of spatial distributions, $1 \,\mathrm{cm}$ grains drift inward 
at a significantly faster speed, producing a different spatial 
distribution.  
The different evolution can be predicted also on the basis of Figure~24 
of \cite{armitage}, showing the radial drift timescale, 
$t_{\mathrm{drift}}$, as a function of the Stokes number. 
Under physical conditions similar to ours ($\alpha \approx 0.001$), 
particles in the size range from $10 \,\mathrm{\mu m}$ to $1 \,\mathrm{mm}$ 
have Stokes numbers ranging from $10^{-5}$ to $10^{-3}$, with a predicted 
$t_{\mathrm{drift}}$ from about $7$ to $3$ times $10^5$ in units of the 
orbital period. The radial drift timescale quickly drops to about 
$2 \times 10^4$ for a Stokes number of $10^{-2}$, corresponding 
to our $1 \,\mathrm{cm}$ size particles. 
The difference in $t_{\mathrm{drift}}$ explains various features in 
Figure~\ref{fig_00}: 
1) the outer edge of the dust disk is mostly
populated by smaller grains, 2) there is an over-density of 
$1 \,\mathrm{cm}$ grains at the outer border of the gap and 
3) the inner region of the gap is partly depleted in large 
grains that have already migrated inside $2$~au (and therefore
removed from the numerical model).

The outcome of this numerical simulation confirms that the common gap in gas leads to the formation 
of a similar gap in the dust distribution and that the two gaps 
approximately overlap at the beginning of the evolution. 
The dust trapped within the gap in horseshoe orbits
is probably short lived because the trajectories
are unstable. However, if the collision rate is
large enough, these particles may accumulate into
larger bodies at Trojan locations before drifting
toward the inner disk.

We acknowledge that this model must be considered
only as indicative of the dust gap formation. 
It is difficult to simulate a more realistic
evolution of the system. In particular the mass
growth of the planets, which is assumed in this
model, occurs on a relatively short timescale. 
In fact, the growth timescale is set by 
disk-limited accretion \citep{lissauer2009}, 
which depends on the thermodynamic state of
the disk and disk-planet tidal interactions. 
In a cold and low-viscosity disk the growth rate
may be significantly longer.

In addition, giant planet cores may 
be massive enough to already form dust gaps prior to the resonant 
trapping \citep{dipierro2017,picogna2015}. 
However, we would not expect that these effects
to significantly change the dynamical evolution 
of the dust. Therefore, for the thermodynamical 
gas conditions used in the model, we would still
expect the formation of a common gap in the dust 
as illustrated in Figure~\ref{fig_00}. 

To explore the later stages of the dust evolution
during the outward migration of the planets, we run
different simulations in which the planets and the
gaseous disk evolve until the resonant state is
established and outward migration ensues.  
Once the planets have migrated outward for some
time, we introduce dust particles in the
simulation, either interior or exterior of the
common gas gap, and compute their evolution over a
long time period. A different behavior is observed 
close to the inner and outer borders of the gas gap
due to the diverse coupling between the gravity and
drag forces acting on the particles. 

The gas gap, shifting as the planets migrate
outward, leads to a progressive radial (outward)
migration of the gap edges. 
At the inner border, gas streaming through the gap
from the outer disk refills the region left empty
by the outward movement of the edge. The dust,
however, unable to efficiently filter through
the planets' orbits, does not replenish the portion
of the disk left devoid of dust by the outward
shift of the inner gap edge.
As a result, the inner border of the dust gap is
left behind while the planets and the gas gap move
outward.

\begin{figure}[hpt]
\centering
\includegraphics[width=1.0\hsize, height=0.9\hsize]{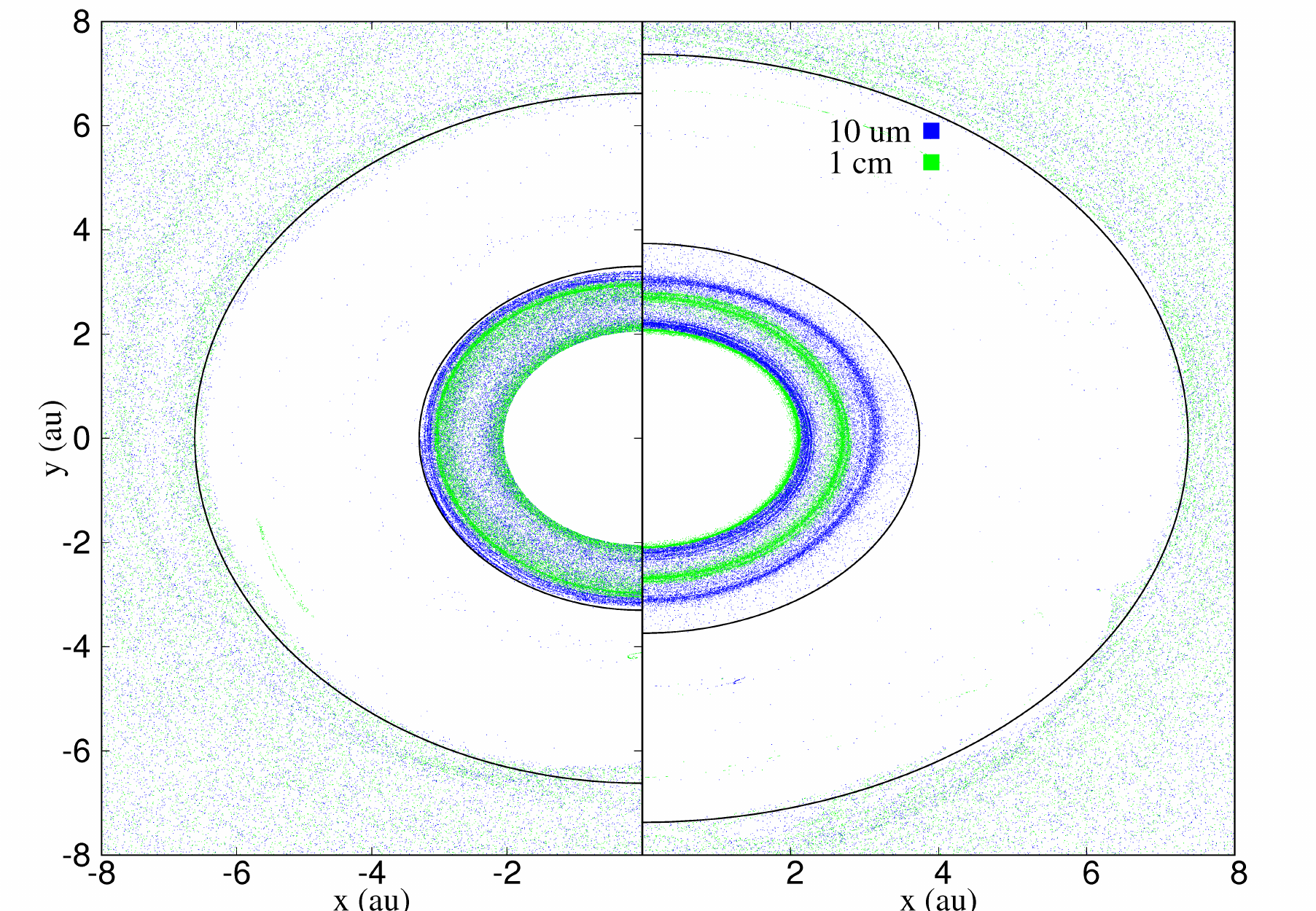}
  \caption{\label{half}  
  Initial dust distribution (left side) and after
  $20$~kyr (right side). The black dashed lines
  show the edges of the common gas gap produced by
  the the planets on resonant orbits. Notice how
  the dust gap broadens as the planets migrate
  outward. Particles of $10 \,\mathrm{\mu m}$ and
  $1 \,\mathrm{cm}$ in radius are plotted. 
  The larger, $1\,\mathrm{cm}$ size grains also
  drift inward at a rate comparable to the outward
  migration rate of the planets, visibly 
  increasing the extent of the corresponding 
  gap in the dust (green dots).}
\end{figure}
This different behavior of the dust and the gas can be observed in 
Figure~\ref{half}. The initial dust distribution on the left side 
evolves in the spatial distribution shown on the right side, after 
about $20$~kyr. 
Also in this case we use the $10 \,\mathrm{\mu m}$ grains 
as representative for the distribution of particles in the range 
$10 \,\mathrm{\mu m}$--$1 \,\mathrm{mm}$, and $1 \,\mathrm{cm}$ 
particles whose drift rate is significantly higher. The gap in the 
dust distribution becomes wider over time and decouples from 
the gas distribution around the gap region. 
The effect is more marked for $1 \,\mathrm{cm}$ size particles since, 
during the outward migration of the gas gap, they also drift 
inward at a comparable rate, increasing the width of the dust gap.  
The dust evolution leading to a broader gap  may complicate 
the interpretation of high resolution observations of dust distributions 
in disks, like those obtained with ALMA. The width of the gap is used to 
estimate the masses of the planets. But in an outward migration scenario 
as that represented here, the dust gap is broader than it would be in the 
static, i.e., non-migrating case. Therefore, to match the observed gap 
width, planets more massive than necessary would be invoked.
It is noteworthy that the orbital eccentricities of
both planets remain below $0.1$, therefore the
observed effect is not due to the increased radial
excursion of the planets \citep[which would result
in eccentric gap edges,][]{dangelo2006}. We also
expect a gap width that depends on the observed
grain size, which may provide an important test to
characterize the type of planetary system
responsible for gaps in the dust distribution.

In the long-term evolution, the strongly reduced
inward dust flux across the planets' orbit and the
radial drift of the dust relative to the gas would
lead to the formation of a cavity, even in absence
of outward migration. The formation timescale,
however, would be relatively long compared to that
due to the planet migration in this resonance.
Also, it would depend on the Stokes number of the
dust \citep[see Figure~24 of][]{armitage}.

An additional effect of the outward migration of
the common gas gap is the accumulation of dust
particles at its outer border. As the planets move
outward, the exterior edge pushes the dust outward,
locally collecting solids and generating an 
over-dense dust region around the exterior border
of the gas gap. Over time, this region of enhanced
density moves outward, carried by the expanding gas
gap edge. As a result, dust concentrations may rise
significantly and attain levels much higher than
those expected from the mere accumulation of
particles due to inward drift driven by gas drag,
since it may occur on a much shorter timescale
(i.e., the migration timescale of the planets).
This is especially the case for small grains,
which have very low (drag-induced) drift velocity
relative to the gas, and observational evidence of
such features may be indicative of a planet
migrating outward.

\begin{figure}[hpt]
  \centering
  \includegraphics[width=1. \hsize]{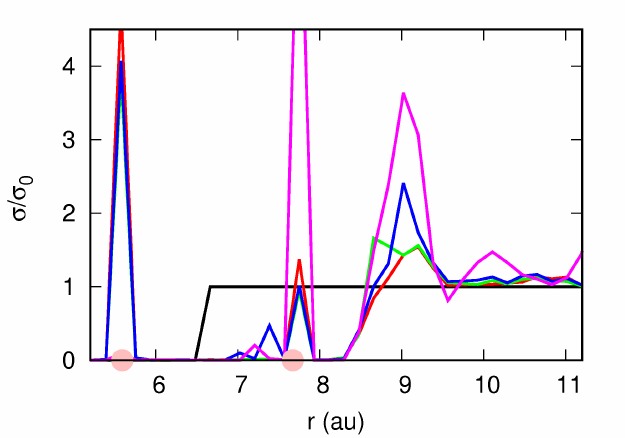}
  \caption{\label{cold32} 
  Histogram of the dust radial distribution after
  $20$~kyr from its inclusion in the simulation. 
  The particles are initially located in the outer
  region of the disk, bordering the gap's exterior 
  edge. Along the $y$-axis, we plot the ratio
  between the initial dust density, $\sigma_{0}$,
  and the final dust density, $\sigma$, 
  as a function of the radial distance. The black
  line marks the initial distribution of dust while
  the red, green, blue and magenta lines indicate
  the distribution after $20$~kyr of 
  $10 \,\mathrm{\mu m}$, $100 \,\mathrm{\mu m}$, 
  $1 \,\mathrm{mm}$, and $1\,\mathrm{cm}$ size
  particles, respectively. 
  As the gas gap moves outward, the dust develops
  an increasing density peak at the exterior
  border, more marked for largest grains due to
  their fast inward drift. The planet positions at
  the end of the simulation are marked by filled 
  pink circles.
 }
\end{figure}

The process of dust accumulation along the exterior
edge of the gap is shown in Figure~\ref{cold32},
where the dust density is derived by grouping the
test particles in radial bins. 
In the Figure, the dust density $\sigma$ is 
normalized to the initial dust distribution,
$\sigma_0$, derived from the positions of the
particles at the beginning of the simulation.
The ratio $\sigma/\sigma_0$ is computed as a
function of the radial distance and it is assumed
that the initial dust-to-gas mass ratio is
constant. 
In this way, we neglect the effects of the frost
line which, however, can be easily accounted for
with a re-normalization of the dust density. 
The distribution of different grain sizes are shown
in different colors and, as expected, for 
$1 \,\mathrm{cm}$ particles the over-density is 
markedly higher due to the contribution of their
rapid inward drift.

The growing density peak at the exterior edge of
the shifting gas gap, which also depends on the
grain size, may provide useful observational
evidence for a pair of planets migrating outward.
If the column density of small dust displays an
anomalously large peak at the outer edge of a gap
observed in a disk (e.g., with ALMA or possibly 
with the Next Generation Very Large Array), this
may indicate a ``sweep-up'' effect of two planets
locked in the 3:2 resonance and migrating outward. 

Just inside the outer edge of the gap where the gas
density acts as a dust trap, two additional large
peaks in the dust density form in the proximity of
each planet, where a significant amount of dust
seems to be collected around them. For this
particular resonance (we will see that this does
not occur in the 2:1 resonance case), the dust
barrier at the exterior edge is not fully efficient
and a non-negligible fraction of dust filters
through the outer edge of the gap during the
coupled migration of the planets. 

\begin{figure}[hpt]
\begin{center}
\includegraphics[width=1.3\hsize,trim={65 0 0 0},clip]{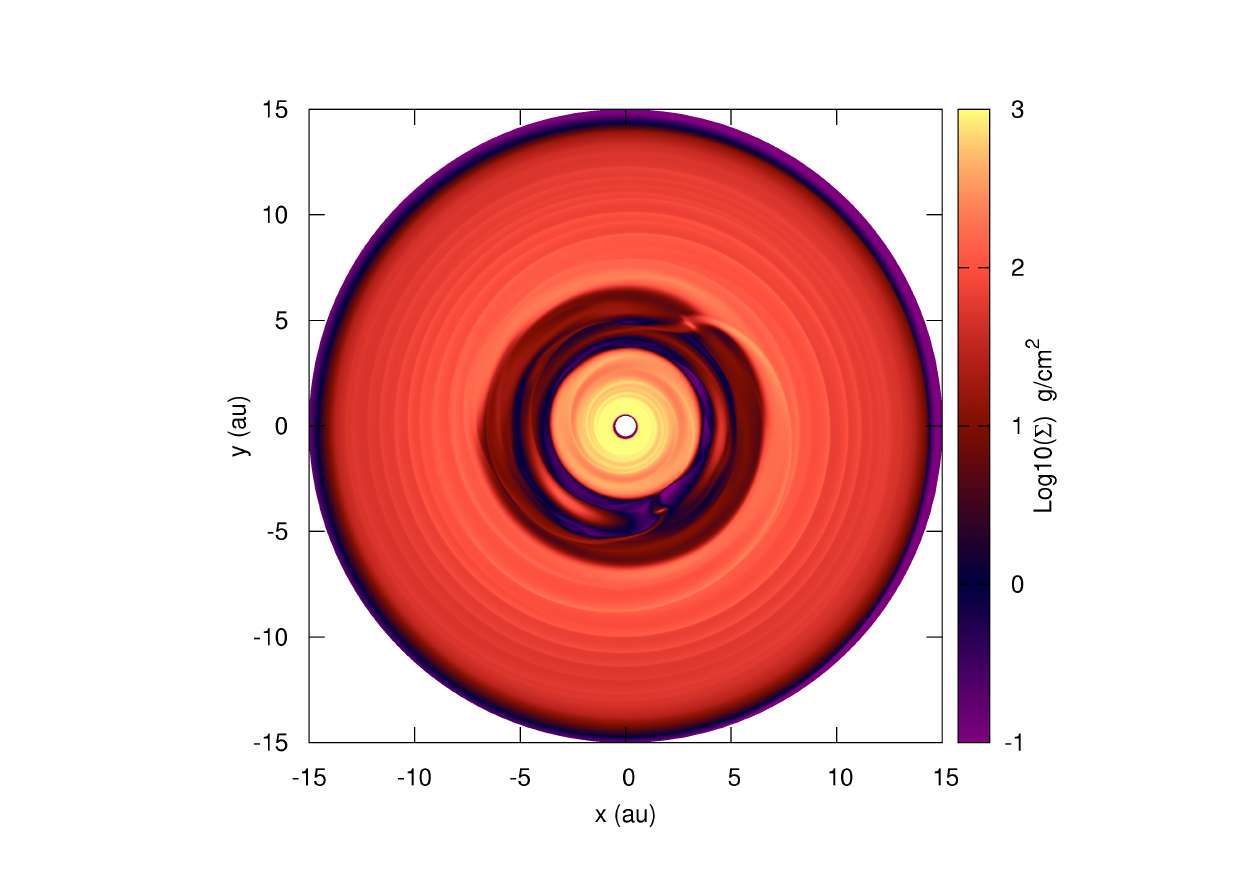}  
\includegraphics[width=1.2\hsize,trim={50 0 0 0},clip]{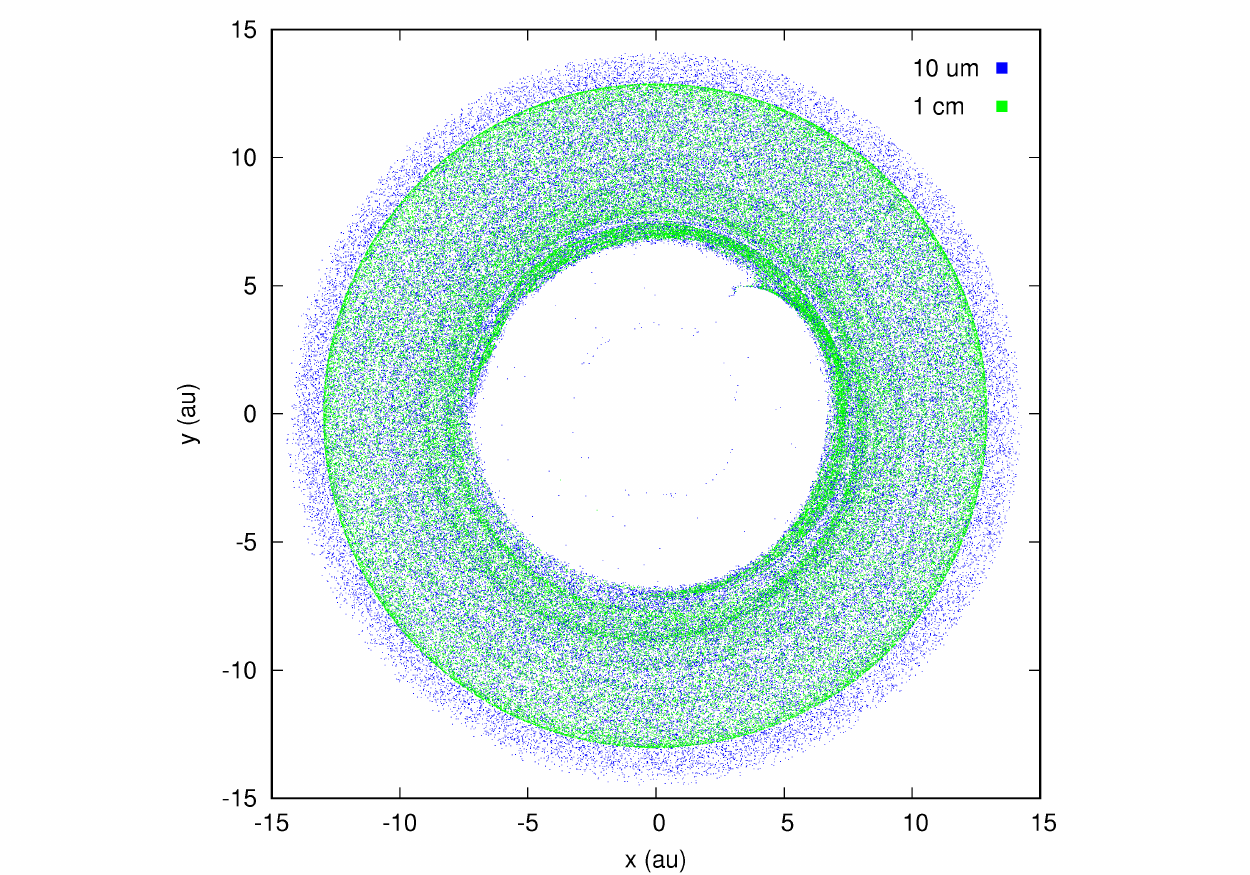}
\end{center}
\caption{\label{32inflow}  
  Gas density distribution (top) and dust spatial
  distribution (bottom) after $10$~kyr from the
  beginning of the simulation. Dust particles are
  entrained within streams crossing the planets'
  orbits and then become trapped around them. }
\end{figure}
This effect can be seen in Figure~\ref{32inflow},
by comparing the distribution of the gas and that
of the dust particles. 
They filter through the gap edge, entrained in the
gas streams, and reach the vicinity of both planets
where they become trapped on bound orbits, and
possibly accrete on the planets. The enhanced 
concentration of dust in the accreted gas may 
impact the growth rate of the planets if it
significantly raises the opacity of the envelope
gas. It is noteworthy that $1 \,\mathrm{cm}$ size
particles filtering through the outer edge of the
gap are almost all captured by the exterior planet
(see Figure~\ref{cold32}) whereas most of the
smaller grains are captured by the interior planet.
This difference is due to the different amount of
coupling between particle and gas dynamics.

\begin{figure}[hpt]
  \centering
  \includegraphics[width=1. \hsize]{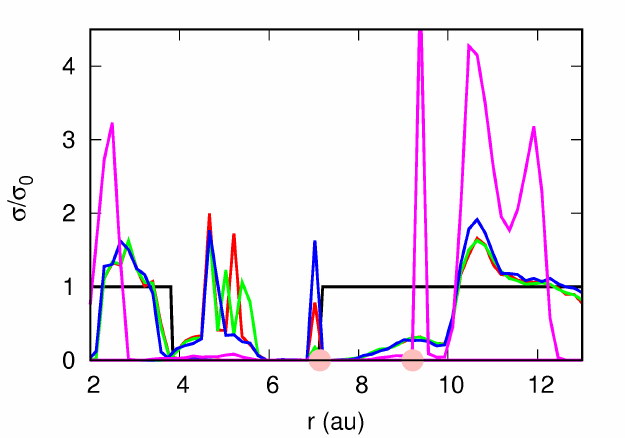}
  \caption{\label{hot32}  
  Histogram of the dust radial distribution after
  $15$~kyr from inclusion in the simulation. The
  particles are initially located in the inner and
  outer region of the disk bordering the edges of 
  the gas gap. As in Figure~\ref{cold32}, the
  y-axis indicates the dust density normalized to
  its initial value, $\sigma_0$, as a function of
  the radial distance. The different lines are as
  in Figure~\ref{cold32}.
  }
\end{figure}

We performed an additional simulation in which the
initial density of the gas is increased by about
$30$\%, 
$\Sigma = 1000 (\mathrm{au}/r)\,\mathrm{g\,cm^{-2}}$,
and the kinematic viscosity is characterized by
$\alpha=10^{-3}$, ten times as high as in the
simulation discussed above. In this scenario, the
outward migration is enhanced (because of the
larger density) and the overall disk
temperature is higher as well. The tidal
perturbations exerted by the planets on the gas are
countered by more vigorous viscous torques, which
reduce their strength and facilitate the transit
of dust through the gap and toward the inner region
of the disk. In Figure~\ref{hot32}, the dust
distribution is shown after about $15$~kyr from the
inclusion of the particles in the model. In the
outer region of the disk, the usual density peak is
observed due to dust accumulation. Smaller
particles from $10 \,\mathrm{\mu m}$ to $1 \,\mathrm{mm}$ in size accumulate at the exterior border but, as in the previous case (see
Figure~\ref{cold32}), a significant fraction of
them filter through the gap edge following the gas
flow. Some are trapped by the interior planet while
most of them cross both orbits and produce an
additional peak around $5$~au. The radial location
of this inner peak is time-dependent and determined
by the drift velocity of the particles. 
The interpretation of the complex features observed
in Figure~\ref{hot32} for the smaller grains may be
problematic since the disk would appear as
separated in three dust rings, which might be
attributed -- erroneously -- to a massive planet
located around $7$~au and by an additional less
massive planet located around $4$~au. In reality,
the planets are located within the main dust gap
and are migrating outward. 

The larger, $1 \,\mathrm{cm}$ grains show a
different behavior compared to the smaller ones. A
significant fraction of particles drift through the 
exterior edge of the gas gap and end up in the
close proximity of the planet, as in the previous
case. Therefore, they do not contribute to the
secondary peak observed for the smaller grains
around $5$~au. At the outer edge, a large 
over-dense region builds up because of the
combination of the outward migration of the gap and 
of the inward radial drift of the grains. An
additional outer peak is present in the radial 
distribution and it is caused by a small increase
of the Stokes number toward the outer regions of
the disk, due to the thermodynamical structure of
the disk. This small increase leads to a faster
drift of the particles, which can overtake the
inner ones producing the smaller peak observed in
the dust distribution around $12$~au. 
This additional feature is not observed in the
distribution of the smaller grains since they drift
inward at a lower speed and their evolution is
mostly dominated by the outward drift of the gas
gap. This different dynamical evolution related to
the size of the particles may represent an
additional test to identify the presence of 
planets in resonance.

\section{The 2:1 resonance}
\label{sec:21mmr}

\begin{figure}[hpt]
  \centering
  \includegraphics[width=1. \hsize]{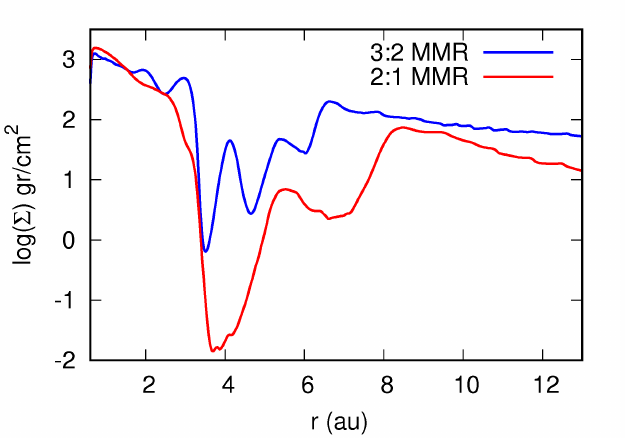}
  \caption{\label{gap}  
  Common gap carved by the planet pair locked in
  the 3:2 (blue) and the 2:1 MMR (red). In the
  latter case the planet separation is larger and 
  their eccentricities are higher, causing the gap
  to be wider.}
\end{figure}
When the power index of the gas density in
Equation~(\ref{eq:Sigma_r}) is equal to $p=3/2$,
the two planets become trapped in a 2:1 MMR because
of the lower gas density at the planets' locations
(see Equation~(\ref{eq:C21})). For the chosen disk
conditions, the pair migrates outward, but at a 
slower speed compared to the 3:2 MMR cases
discussed in the previous section. The resonant 
forcing, however, has a stronger effect on the
orbital eccentricities of the planets than in
the previous cases (see Figure~\ref{migra}). The
wider separation between the planets, combined to
their higher orbital eccentricity, leads to a gas
gap for the 2:1 MMR case that is broader than that
of the 3:2 MMR, as illustrated in Figure~\ref{gap}.  

The gap also appears deeper than in the 3:2 MMR
case but this may be due to the different
evolutionary times of the two configurations shown
in the Figure. Since the speed of the outward 
migration for the 3:2 MMR case is faster than in
the 2:1 MMR case, to compare the gaps at similar
locations in the disk different evolutionary 
times were chosen for the two cases illustrated in
Figure~\ref{gap}. The gap in the 2:1 MMR is
significantly more evolved, which is consistent
with its greater depth. The lower migration speed
of the 2:1 MMR case also contributes to make the
gas gap deeper.

\begin{figure}[hpt]
  \centering
  \includegraphics[width=1. \hsize]{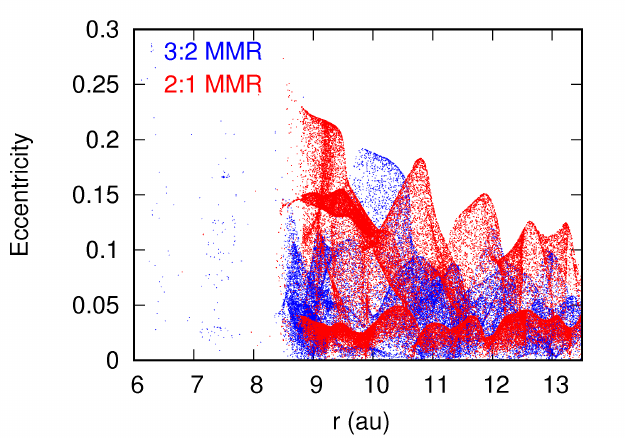}
  \caption{\label{ecce_dust}  
  Distribution of the dust particle eccentricities
  close to the outer edge of the gap for the 3:2
  (blue) and 2:1 MMR (red),
  respectively. The evolutionary times are
  different for the two resonances and are selected
  in order to have the gap outer edge at similar
  locations and to compare the grain eccentricity
  values at similar distances from the star.}
\end{figure}
The higher eccentricity of the planets affects not
only the morphology of the gaseous disk but also
the orbital parameters of the dust grains. 
In Figure~\ref{ecce_dust}, the eccentricities of
the dust particles in the 3:2 and 2:1 resonance
are compared at different evolutionary times 
during the planets' outward migration, when the
exterior border of the gaps in the dust are
approximately equal. In the configuration shown,
the planets are at different radii (in the two
simulations), but the radial locations 
where dust is collected are similar in the two
cases because the gap generated by the 2:1 MMR
locking is broader. Additionally, in the case of 
the 2:1 MMR, the dust grains move on more eccentric
orbits compared to the 3:2 MMR case, in particular
close to the gap edge. 

\begin{figure}[hpt]
  \centering
  \includegraphics[width=1.3 \hsize,trim={100 0 0 0},clip]{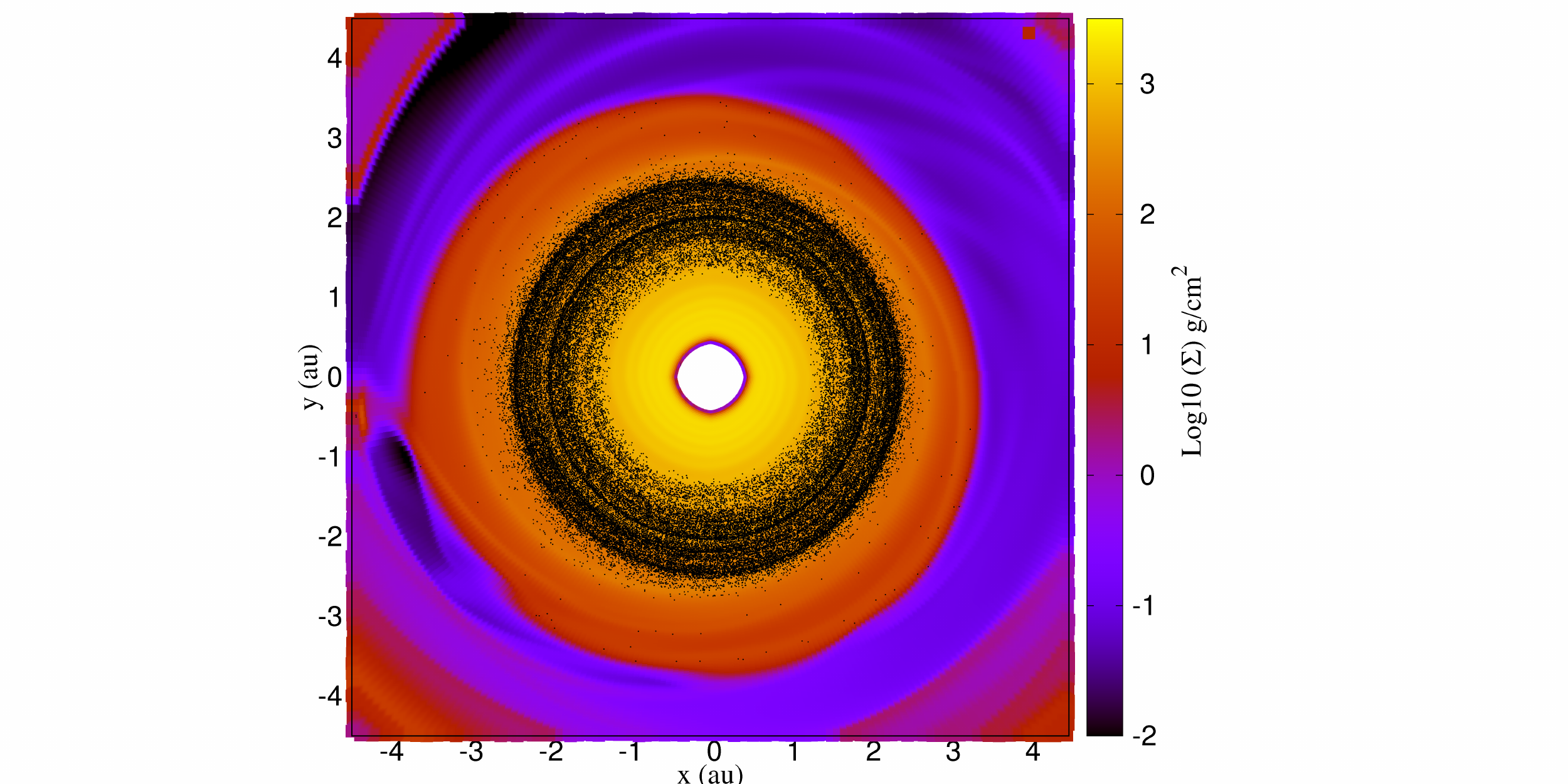}
  \caption{\label{dust_gas} 
  Gas density distribution and dust particles 
  (only those of $10 \,\mathrm{\mu m}$ in radius)
  in the inner region of the disk after $30$~kyr 
  of outward migration of the planets, locked in
  the 2:1 MMR. The dust particles are marked by
  black dots.}
\end{figure}
Notwithstanding the different shapes of the gaps
and eccentricity of the particles, the same 
phenomenon observed in the gap region of the
planets in the 3:2 MMR takes place in the 2:1 MMR
case. The dust particles populating the inner disk
are not replenished as the planets migrate outward,
so that the inner edge of the dust gap appears to
recede from that of the gas. However, this happens
on a longer timescale compared to the 3:2 MMR case,
since the outward migration of the planets is 
slower. Also, in the 2:1 MMR case, the gap in the
dust progressively broadens as the planets migrate
outward while that of the gas remains approximately 
constant (over the same timescale). This effect is
clearly shown in Figure~\ref{dust_gas}, where the
dust particles are plotted against the gas density
distribution. After $20$~kyr from the inclusion of 
the dust grains in the calculations, the gap has 
significantly moved outward and the gas has
refilled the volume left empty by the outward
movement of the inner planet. The dust, however, 
behaves differently and the initial ring does not
diffuse outward. 
On longer timescales, we expect the size of the gap
in the dust to become progressively wider compared
to that in the gas, an effect that may be detected
by observations. For example, with ALMA the spatial
distribution of sub-millimeter emission from solids 
can be compared to CO observations of gas to test
for possible differences in the gap sizes. If some
discrepancy is detected and there is an enhanced
dust emission around the outer gap edge, 
it may be argued that the overall gap in the disk
is due to two large planets in MMR and migrating
outward.  

\begin{figure}[hpt]
  \centering
  \includegraphics[width=1. \hsize]{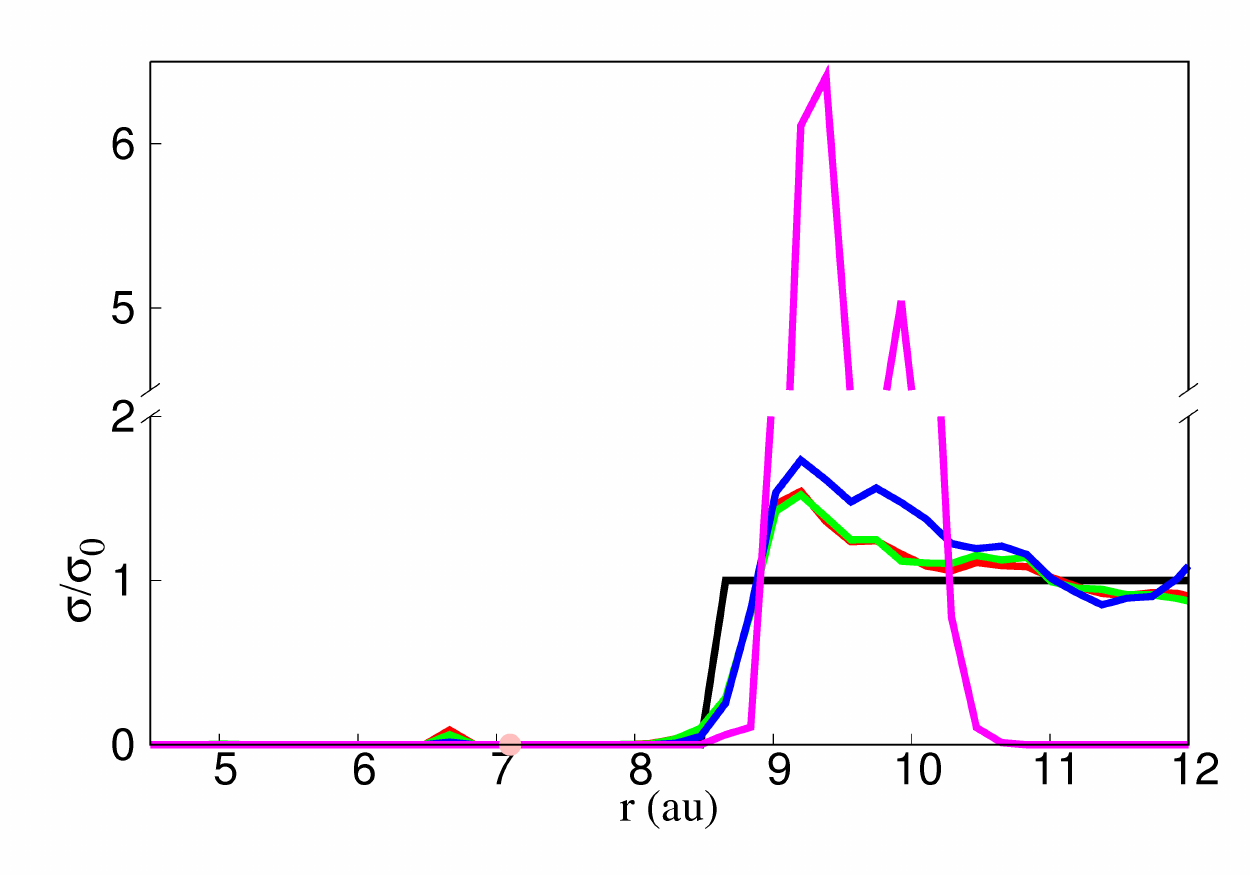}
  \caption{\label{histo2}  
  Histogram showing the normalized distribution 
  of dust particles after $30$~kyr of evolution
  within the disk. 
  Dust accumulates at the exterior border of the 
  gas gap, creating a dense dust ring, which can 
  be detected by observations. Its density is
  significantly higher compared to the case of the
  3:2 MMR (for similar evolution times), 
  taking also into account the slower outward
  migration rate of the planets. }
\end{figure}
In the case of the 2:1 MMR, the dust over-density
that forms at the outer border of the gap (see
Figure~\ref{histo2}) builds up
at a lower rate because of the slower outward
migration speed of the gas gap compared to that of
the 3:2 resonance case (see Figure~\ref{migra}).
The outer edge of the gap is also very efficient at
blocking dust grains of all sizes and only a very
small fraction of the dust (of any size) crosses
the outer border of the resonant gap. Almost no
particle is found close to the planets, contrary to
the 3:2 MMR case, in which a non-negligible amount
of dust was dragged toward the planets (compare
Figure~\ref{histo2} with Figures~\ref{cold32} 
and \ref{hot32}). This
suggests that at later times the density peak at
the outer edge of the gas gap will be more
prominent compared to that observed in the 3:2 MMR
case. This is already partly evident for the
larger $1 \,\mathrm{cm}$ size particles, which
accumulate at the outer edge and build up a high
density ring during the course of the calculation
(see Figure~\ref{histo2}).
The continuous supply of additional dust coming
from the outer disk regions (which are not 
included in our model) will lead to a significant
growth of the density over time. This different
behavior between the two MMR configurations may
provide an opportunity to distinguish between 
them from observations. 

\begin{figure}[hpt]
   \includegraphics[width=\hsize]{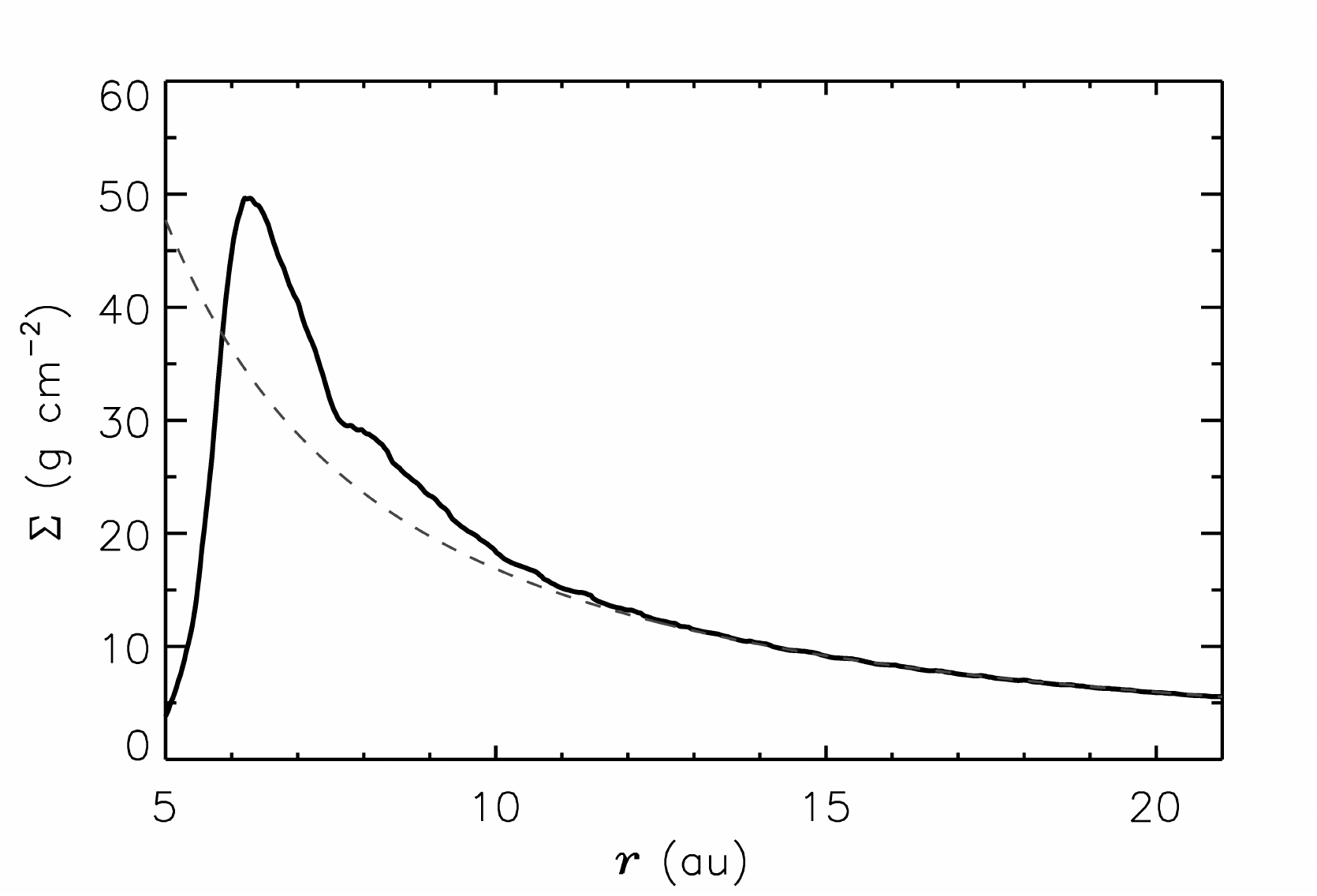}
   \includegraphics[width=\hsize]{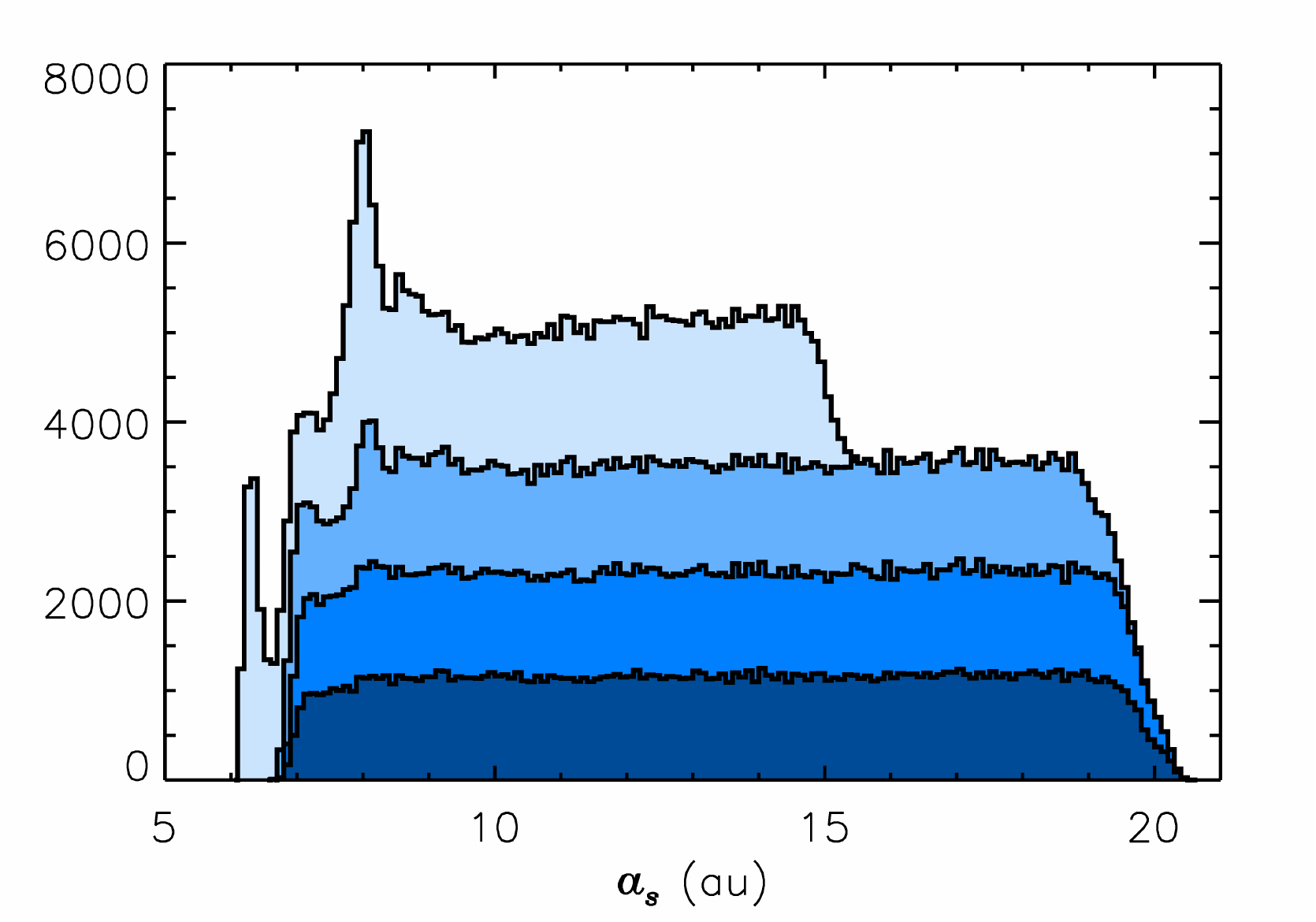}
   \caption{
   Top: Azimuthally averaged gas surface density
   (solid line) exterior of a Jupiter-Saturn pair,
   locked in the 2:1 MMR and migrating outward 
   (see Figure~\ref{ae21}). 
   The dashed line represents the initial gas
   density distribution.
   Bottom: Cumulative histograms of the semi-major
   axes of dust particles, $a_{s}$, of radii
   $10\,\mathrm{\mu m}$, $100\,\mathrm{\mu m}$,
   $1\,\mathrm{mm}$, and $1\,\mathrm{cm}$ 
   (darker to lighter color shades, respectively).
   Note that the evolution of the grains,
   $\approx 2$~kyr, is much shorter than that in 
   Figure~\ref{histo2}.
            }
   \label{sig_as}
\end{figure}
In Figure~\ref{sig_as}, we show results from the
additional locally isothermal calculation discussed
in Section~\ref{sec:numerical} (see also
Figure~\ref{ae21}). The pair of planets is locked
in the 2:1 MMR and is migrating outward (with 
Jupiter and Saturn at around $2.5$ and 
$3.97\,\mathrm{au}$, respectively, at the evolution 
time of Figure~\ref{sig_as}). The gas surface
density (top panel) is displayed starting from the
outer edge of the common gas gap. The dust is
initially deployed from $7$ to $20\,\mathrm{au}$,
uniformly in radial distance and in equal numbers
per size bin. The Stokes numbers at deployment
range from $\approx 10^{-4}$ (for the smallest
particles) to several times $10^{-2}$ 
(for the largest particles).
The bottom panel shows the cumulative histogram of
the semi-major axis ($a_{s}$) of particles of radii
$10\,\mathrm{\mu m}$, $100\,\mathrm{\mu m}$,
$1\,\mathrm{mm}$, and $1\,\mathrm{cm}$, indicated
respectively by darker to lighter colors. Each
colored region (i.e., the region in between two
neighboring black curves) provides the actual
number of particles as a function of $a_{s}$. The
largest particles drift inward the fastest, as
mentioned above. Inside of $6\,\mathrm{au}$, 
the steep gas density gradient (see top panel)
should efficiently push the dust outward,
preventing it from reaching the interior disk.
In fact, barely any particles is able to filter 
through the gas gap edge toward the planets.

The dust evolution shown in the bottom panel of
Figure~\ref{sig_as}, $\approx 2$~kyr, is
significantly shorter than that discussed above
and illustrated in Figure~\ref{histo2}, hence the
lower relative peaks at the exterior edge of the
gas gap. The formation of dense dust rings is
also delayed by the lower $\Sigma$ of this model
compared to that of the model represented in
Figure~\ref{histo2}. However, as time progresses,
the density peak in the dust distribution of
$1\,\mathrm{cm}$ grains is expected to increase.
The local density of the smaller particles is also
expected to increase, but on the longer timescale
of the planets' outward migration
(see Figure~\ref{ae21}).

\section{Discussion and Conclusions}
\label{sec:end}

The capture of two planets in resonance may be a
common event in the early phases of evolution of
planetary systems, when the circumstellar gas disk
is still present. Convergent migration may drive an
outer, less massive planet close to an inner more
massive planet until capture occurs in either the 
2:1 or the 3:2 MMR. The type of resonance and the
subsequent evolution of the planets depend on many
parameters, like the disk's surface density, 
the planet masses, and the disk viscosity. 
We focus here on a scenario where the planets, once
trapped in resonance, migrate outward. 
This mechanism is invoked, for example, in the 
grand tack scenario to explain some of the
properties of the inner solar system.

In this paper we outline some peculiar traits of the 
dust spatial distribution produced by the perturbations of the 
resonant planets. These traits
may produce detectable signatures which 
can reveal the presence of planets in 
high resolution images
taken with instruments like ALMA or SPHERE (or with future
instruments such as the Next Generation Very Large
Array).


Through numerical models of the evolution of gas,
dust, and a pair of planets with two different
hydrodynamical codes, we first show that the common
gas gap induces an extended gap also in the dust. This
creates an ambiguity when interpreting a dust gap
observed in a circumstellar disk, because the gap
width can be modeled with either a single massive
planet or a pair of smaller planets in resonance.
However, the possible outward migration of a pair
of planets locked in MMR induce some
characteristic features in the dust distribution
that may be used to resolve this ambiguity. 

During the outward migration, the dust gap becomes
progressively wider compared to the gas gap since
the grains populating the region inside the inner
edge of the gas gap do not move outward to refill
the region behind the migrating planets. 
This broadening depends on the migration rate and
occurs on a shorter timescale in the 3:2 MMR case
compared to the 2:1 MMR case. Different disk
conditions may alter this result, but capture in
the 2:1 MMR followed by outward migration requires
a cold and low-viscosity disk.
In addition, for the 2:1 MMR, just after capture,
the eccentricity begins to grow, leading to a 
progressive initial broadening of the gas gap while
it moves outward. However, even if over a longer
timescale, the dust gap formed by a pair of planets
locked in the 2:1 MMR eventually decouples from 
the gas gap. This behavior of the dust has two
important implications. First, the inner edge 
of the dust gap marks the radial location at which
the planets reversed the direction of their
migration (assuming that inward drift of the grains
due to gas drag occurs over much longer
timescales). Based on estimates gap sizes, 
it should be possible to deduce where the planets
became locked into resonance. Secondly, in the case
of a prolonged outward migration, the dust gap
would become progressively wider while that in 
the gas would approximately maintain its shape and
move with the planets. Therefore, if the dust 
gap significantly differs from the gas gap and it
is much wider, then it may be attributed to a pair
of planets migrating away from the star rather than
to a single (non-migrating) massive planet. 

An additional feature characteristic of the outward
migration of two planets locked in resonance is the
formation of a high and quickly growing dust
density peak at the exterior edge of the gas gap.
The outward drift of the gap and the ability to
trap dust act in synergy, gradually accumulating
grains in a dense layer at the outer edge of the
gap. This is particularly relevant for large size
grains ($\sim 1 \,\mathrm{cm}$, Stokes numbers
$\sim 10^{2}$) whose fast inward drift contributes 
to collect particles at the outer edge of the gap. 
Also for this effect, there are significant
differences between the two resonances. 

In the case of the 2:1 MMR, the exterior edge of
the gas gap is a very effective barrier and dust
grains are unable to cross it. Consequently, the
radial drift of the gas gap pushes the dust outward
and produces a significant build-up of dust, which
would appear as a bright ring in high-resolution
images. 
In the case of the 3:2 MMR, the faster outward
migration of the planets would lead to a quicker
dust accumulation at the exterior edge of the gap
but the dust barrier effect for this resonance is
less effective and a significant fraction 
of dust particles can cross the gap and stream 
toward the planets, but without reaching the inner
disk region. It is then expected that, for this 
resonance, the outer dust ring would be less
prominent (in observations) compared to that 
produced by the 2:1 resonance. 
If the gas density, and hence the outward migration
rate, increases then an additional peak in the dust
appears, moving in from the outside and rendering
the disk ring structure more complex.
This additional feature is present only in the
distribution of dust particles from 
$10 \,\mathrm{\mu m}$ to $1 \,\mathrm{mm}$ in 
size. In fact, almost all $1 \,\mathrm{cm}$ grains
tend to be collected around the exterior planet, 
once they filter through the outer edge of the
gas gap.

In light of these results, what can be argued when 
a gap is observed in a circumstellar disk? It may
be due to a single planet within a range of masses
or it can be ascribed to two less massive planets
locked in resonance. 
If they are migrating outward, then we expect two
distinctive features: a decoupling between the dust
and gas gap and an over-dense layer at the 
exterior edge of the gas gap. An interesting
question concerns the case of a pair of planets in
MMR and migrating inward. Should we expect, in this 
case too, a decoupling between the gas and dust gap
with the inverted roles of the dust gap edges? 
The situation is not symmetric. During outward
migration, dust particles at the the interior gap
edge would drift inward under the action of gas
drag (although on a timescale longer than the
migration timescale), contributing to the
broadening of the dust gap. In the case of inward
migration, the dust at the exterior edge would move
in the same direction as the gas gap, and
the decoupling effect between the dust and gas
would be possibly less conspicuous.
A numerical investigation is required to get more
definitive answers for some reasonable ranges of
disk parameters. 

The effect of gas turbulence on the solids was not considered in these calculations \citep[see, e.g., discussion in][]{paamel2006,zhu2012}. Turbulence transport may induce diffusion of particles across the gap edges, possibly reducing the local density of the solids exterior of Saturn's orbit. The consequences, however, are difficult to quantify without direct simulations (and appropriate assumptions on the nature of turbulence). In fact, more generally, gas turbulence can also impact the properties of the gap edges.

An additional effect, neglected in this study, is
the back-reaction of the dust particles onto the
gas, which may be important when the dust-to-gas
ratio grows in the proximity of the exterior 
border of the gap. The back-reaction might alter
the morphology of the gap edge, affecting the dust
dynamics in this region and allowing particles to
filter through. 
However, according to \citep{Taki2016}, this may 
occur only when the dust-to-gas ratio approaches
$\approx 1$ and the restoration process of
the pressure maximum is slower than the deformation
process (by the dust back-reaction). This may not
be the case of a pressure maximum created by the
competing actions of tidal forces exerted by the
planets and the viscous forces exerted by the gas.
This is a point, however, deserves a dedicated
investigation. It is noteworthy that the high 
dust density build-up at the exterior border of the
gas gap may also favor accumulation and growth 
processes. The size distribution of the solids in
this region may then be different from that of the
surrounding disk, providing an additional
observational test for the presence of two planets
in resonance.

\begin{acknowledgements}
We thank an anonymous reviewer for helpful comments.
G.D.\ acknowledges support from NASA's Research
Opportunities in Space and Earth Science (ROSES).
G.P.\ acknowledges support from the DFG Research Unit ``Transition Disks'' (FOR~2634/1, ER~685/8-1).
Resources supporting the work presented herein were
provided by the NASA High-End Computing (HEC)
Program through the NASA Advanced Supercomputing
(NAS) Division at Ames Research Center.
\end{acknowledgements}

\bibliographystyle{aasjournal}



\end{document}